\documentclass[11pt,a4paper]{article}
\pdfoutput=1  
\usepackage{amsmath,amsfonts,amssymb,amsbsy}
\usepackage{graphicx}
\usepackage{color}
\usepackage{booktabs}
\usepackage{multirow}
\usepackage{multicol}
\usepackage{subfigure}
\usepackage{alpha}
\usepackage{macros_alpha}
\providecommand*{\ord}{\ensuremath{\text{O}}}

\providecommand*{\e}{\ensuremath{\text{e}}}
\providecommand*{\De}{\ensuremath{\mathcal{D}}}
\providecommand*{\psibar}{\ensuremath{\overline{\psi}}}

\renewcommand{\strange}{\mathrm{strange}}

\begin{document}

\preprintno{%
CP3-Origins-2016-032 DNRF90\\
}

\title{%
On reweighting for twisted boundary conditions 
}

\author[ode]{Andrea~Bussone}
\author[ode]{Michele~Della~Morte}
\author[ode]{Martin~Hansen}
\author[ode]{Claudio~Pica}

\address[ode]{CP$^3$-Origins, University of Southern Denmark, Campusvej 55, 5230 Odense M, Denmark}

\begin{abstract}
We consider the possibility of using reweighting techniques in order to correct
for the breaking of unitarity when twisted boundary conditions 
are imposed on valence fermions in simulations of lattice gauge theories.
We start by studying the properties of reweighting factors and their variances at tree-level.
That leads us to the introduction of a factorization for the fermionic reweighting determinant.
In the numerical, stochastic, implementation of the method, we find that 
the effect of reweighting is negligible in the case of large volumes
but it is sizeable when the volumes are small and the twisting angles are large. 
More importantly, we find that for un-improved Wilson fermions, and in small volumes, 
the dependence of the critical quark mass on the twisting angle
is quite pronounced and results in large violations of the continuum dispersion relation. 
\end{abstract}


\maketitle 

\section{Introduction}

Simulations of field theories on discretized, Euclidean, lattices are necessarily performed
in a finite space-time volume. That requires imposing boundary conditions on the fields and 
although different choices have the correct infinite volume limit, the way that is approached
depends on the particular choice (see for example the discussion in~\cite{Lucini:2015hfa,Borsanyi:2014jba} for the case
of lattice QED). The actual setup may  affect not only the physical
results by finite volume effects but also the algorithmic efficiency and sampling properties 
of the simulations. Recent examples of the latter can be found in~\cite{Luscher:2011kk} for the use of open boundary conditions in lattice QCD
to bypass the freezing of topology, and in~\cite{DellaMorte:2010yp} for the use of generalized boundary conditions in order to exponentially
improve the signal to noise ratio in glueball correlation functions computed in the pure gauge theory. 
When considering lattice QCD, (anti)-periodic boundary conditions in (time)-space are usually imposed on the fermionic fields.
That leads to a quantization of the spatial momenta in units of $2\pi/L$ with $L$ the spatial extent of the lattice.
For the lowest non-zero momentum to be around 100 MeV, lattices of about 12 fm extension are hence needed.
That is still very demanding from the computational point of view if at the same time one wants
to keep discretization effects under control, which typically requires considering lattice spacings $a$ of about 0.1 fm and below.
In addition, for several applications relevant for phenomenology it is desirable not only to reach small
momenta, but also to have a fine resolution of them. Examples include form factors, as those describing
the $K \to \pi \ell \nu$ transition, the charge radius of the pion, and the hadronic vacuum polarization of the photon, relevant for the
muon $g-2$ anomaly.

Twisted boundary conditions, first introduced in~\cite{deDivitiis:2004kq,Bedaque:2004kc} offer a way to continuously vary momenta in lattice QCD,
and have indeed been used for all the computations mentioned above~\cite{Boyle:2007wg,Brandt:2013dua,DellaMorte:2011aa}.
They have also been applied to the calculation of renormalization factors in the RI-MOM scheme~\cite{Arthur:2010ht} and to the
matching between Heavy Quark Effective Theory and QCD using correlators defined in the Schr\"odinger Functional~\cite{DellaMorte:2013ega}.
Twisting amounts to imposing periodic boundary conditions up to a phase (the twisting angle $\theta$) for fermions in the spatial directions.
In actual simulations the partially twisted setup is usually adopted, where twisting is only applied in the valence sector whereas fermions
in the sea are kept periodic. That introduces a breaking of unitarity as a boundary effect, which therefore 
is expected to disappear in the infinite volume limit, as it has explicitly been checked using Chiral Perturbation Theory in~\cite{Sachrajda:2004mi}. 

This suggests that reweighting techniques as those employed in~\cite{Finkenrath:2013soa} for the case of mass-reweighting
could be used here in order to change the periodicity conditions for fermions in the sea.
We will see in the following that the resulting reweighting factors are ratios
of fermionic determinants, which tend to the value one in the infinite volume limit.
Therefore, as mentioned above, if any effect of unitarity violations can be seen, then
that is expected  to happen in rather small volumes, where the reweighting factors
(which are extensive quantities) can be reliably computed and used as correction factors 
if needed.

A preliminary account of the present studies appeared in~\cite{Bussone:2015yja}.
The paper is organized as follows; in Section~2 we collect definitions and details on the setup we used and in Section~3
we present exact results obtained at tree-level. Those will turn out to be useful in optimizing 
the numerical techniques employed for the evaluation of the reweighting factors and their variances.
Simulation parameters and Monte-Carlo results
are presented in Section~4.
The pion and quark mass dependence on the twisting angle are discussed in Section~5.
Section~6 contains our conclusions.

\section{Definitions and setup}

The generic boundary conditions for matter fields in lattice QCD formulated on a torus are nicely discussed 
in~\cite{Sachrajda:2004mi}. There it is pointed out that it is sufficient to require that the action is single valued on
the torus, whereas the fields themselves do not need to be. Periodicity conditions on the fermions can therefore be of the form
\color{black}
\begin{equation}
	\Psi\left(x+L_\mu\hat{\mu}\right) = V_\mu \Psi\left(x\right)\;, \quad \mu=1,2,3 \,,
\end{equation}
where $\Psi$ is a flavor multiplet and $V_\mu$ represents a unitary transformation
associated to a symmetry of the action.
Similarly, for the $\overline{\Psi}$ field one requires
\begin{equation}
\overline{\Psi}\left(x+L_\mu\hat{\mu}\right) =  \overline{\Psi}\left(x\right) V_\mu^\dagger\;, \quad \mu=1,2,3 \,.
\end{equation}
Considering now generic values of the diagonal quark mass matrix, one concludes that  $V_\mu$  also has to be diagonal in flavor space, i.e
\begin{equation}
	\psi\left(x+L_\mu\hat{\mu}\right) = e^{i\theta_\mu} \psi\left(x\right)\;, \quad \mu=1,2,3 \,,
\label{theperiodicity}
\end{equation}
where the twisting angles $\theta_\mu \in [0,2\pi)$ have been introduced for each flavor and $\psi$ is now 
one component of the $\Psi$ multiplet.
\color{black}

Equivalently, one can fix the fermionic fields to be periodic and 
introduce a constant U$(1)$ interaction with vanishing electric and magnetic fields, vanishing electric potential but constant vector potential \cite{Bedaque:2004kc,Sachrajda:2004mi}. In lattice gauge theories, such an interaction is implemented by transforming the standard QCD links $U_\mu(x)$ in the following way
(setting $a=1$)
\begin{equation}
\label{eq:modified_links}
					\tilde{U}_\mu(x)= \begin{cases}
						\e^{i\theta_\mu/L_\mu}U_\mu(x)\;, \quad \mu=1,2,3\\
						U_0(x)\;, \;\;\,\quad\qquad \mu=0
					\end{cases}.
\end{equation}
In order to see the equivalence, it is enough to observe that the phase can be re-absorbed by re-writing the, now periodic, $\psi$ fields in terms of
\begin{equation}
\label{eq:modified_fermion}
					\tilde{\psi}(x)=
						\e^{i\, \overrightarrow{\left(\frac{\theta}{L}\right)} \, \vec{x}}\psi(x)\;,
\end{equation}
and to notice that the $\tilde{\psi}$ are indeed periodic up to a phase, as for Eq.~\ref{theperiodicity}.
The spatial Fourier modes of the $\langle  \tilde{\psi}(x)   \overline{\tilde{\psi}}(0) \rangle$ propagator are of
the shifted form $e^{i\left(\vec{k}+\frac{\vec{\theta}}{L}\right)\vec{x}}$ with each component of the vector $\vec{k}$ being an integer multiple
of $2\pi/L$ (for the special but rather typical case $L_1=L_2=L_3=L$).
In this sense twisting allows to continuously vary momenta as mentioned in the introduction.
The corresponding amplitudes can be extracted from the Fourier decomposition of 
the propagator $\langle  \psi(x)   \overline{\psi}(0) \rangle$, which satisfies
periodic boundary conditions. In practice, $\rm SU(N_c)$ gauge configurations are typically produced 
for one specific choice of $\theta$, and the angle is then varied only when computing the
quark propagators, which is cheaper in terms of CPU-time with respect to the generation
of configurations. As a consequence, the quark propagators in the sea and valence sectors differ,
which causes a breaking of unitarity already at the perturbative level.
This effect however can be studied in a rather straightforward way, as done here. 

In the following we will use the un-improved Wilson action, with the links $U_\mu(x)$ replaced by the $\tilde{U}_\mu(x)$ 
as in Eq.~\ref{eq:modified_links}. That replacement clearly does not affect the plaquettes and therefore the pure gauge term in the action.
Only the covariant derivatives and the Wilson term are modified. Hence, once the fermionic degrees of freedom are integrated
out on each $\rm SU(N_c)$ gauge background, the $\theta$-dependence from the sea sector is completely absorbed in the fermionic determinant.

In order to apply reweighting techniques, let us imagine we want to compute the value of some observables
for one choice of bare parameters  $B = \{\beta', m'_1, m'_2, \dots, m'_{n_f}, \theta'_\mu, \dots\}$, using
the configurations produced at a slightly different set of parameters  $A = \{\beta, m_1, m_2, \\ \dots, m_{n_f}, \theta_\mu, \dots\}$.
To this end, one needs to compute on each configuration of the $A$-ensemble the reweighting factor $W_{A B} = P_B /P_A$, 
which is the ratio of the two probability distributions and it is an extensive quantity, $P_A[U] = \e^{-\text{S}_{\text{G}}[\beta, U]} \prod_{i=1}^{n_f}\det\left(D[U,\theta]+m_i\right)$. In the last expression  we have explicitly indicated the dependence of the Dirac operator on the twisting angle.
The expectation values on the $B$-ensemble can then be expressed as
\begin{equation}
\langle\mathcal{O}\rangle_B = \frac{\langle\widetilde{\mathcal{O}} W_{A B}\rangle_A}{\langle W_{A B}\rangle_A}\, ,
\end{equation}
with $\widetilde{\mathcal{O}}$ being the observable defined after Wick contractions, 
and $\langle\dots\rangle_{A}$ indicates that expectation values have to be taken on the  $A$-ensemble.
Specializing to the case where
only the periodicity angles of the fermionic boundary conditions are changed from one bare set to the other, 
we obtain the following expression of the reweighting factor 
\begin{align}
		W_\theta = \det\left(D_W[U,\theta]D_W^{-1}[U,0]\right) = \det\left(D_W[\tilde{U},0]D_W^{-1}[U,0]\right),
\end{align}
where we have also chosen $D$ to be $D_W$, i.e., the massive Wilson Dirac operator.
Under certain conditions ratios of determinants as those above can be estimated stochastically.
In general, for a normal matrix $M$ whose eigenvalues have positive real parts, the following representation of the determinant can be used~\cite{Finkenrath:2013soa}
\begin{equation}
\label{eq:appl_cond}
\frac{1}{\det M}  = \int \De\left[\eta\right] \exp\left(-\eta^\dagger M \eta\right)<\infty\, \Longleftrightarrow\, \mathbb{R}\text{e}\lambda\left( M\right)>0.
\end{equation}
The positivity condition ensures that the integral converges. The expression can clearly be evaluated stochastically.
The distribution  $p(\eta)$ of the vectors $\eta$ is usually taken to be gaussian, and in that case, the determinant (or its inverse) can be written as 
\begin{align}
\frac{1}{\det M}  = \bigg\langle \frac{\e^{-\eta^\dagger M \eta}}{p(\eta)} \bigg\rangle_{p(\eta)} = \frac{1}{N_\eta} \sum_{k=0}^{N_\eta} \e^{-\eta_k^\dagger (M-\mathbf{1}) \eta_k} + \ord\left(\frac{1}{\sqrt{N_\eta}}\right).
\end{align}
It is straightforward to generalize the positivity condition above in order to ensure the convergence of the stochastic estimates of all gaussian moments.
In the case of an hermitean matrix one obtains
\begin{align*}
\bigg\langle \frac{ \e^{-2\eta^\dagger M \eta} }{p(\eta)^2} \bigg\rangle_{p(\eta)} & = \int \De\left[\eta\right] \exp\left[-\eta^\dagger (2M-\mathbf{1}) \eta\right] < \infty\, \Longleftrightarrow\, \lambda\left( M\right)>\frac{1}{2},\\
\vdots \hspace{1.4cm}&\\
\bigg\langle \frac{ \e^{- N \eta^\dagger M \eta} }{p(\eta)^N} \bigg\rangle_{p(\eta)} & = \int \De\left[\eta\right] \exp\left[-\eta^\dagger [NM-(N-1)\mathbf{1}] \eta\right] <\infty \Longleftrightarrow\, \! \lambda\left( M\right) \! >\! \frac{N-1}{N}\! \underset{N\rightarrow\infty}{\longrightarrow} \!1.
	\end{align*}
All eigenvalues should therefore be larger than unity. In particular, in the numerical studies presented here we will
always consider the square of the hermitean version of the Wilson Dirac operator $Q= \gamma_5 D_W$, which is to say
we consider the case of two degenerate flavors.

\section{Tree-level studies}

\begin{figure}[h!t]
\begin{center}
\subfigure[\emph{Mean of the reweighting factor at tree-level for $\theta = 0.1$ as a function of $L$.}]{\includegraphics[scale=0.49]{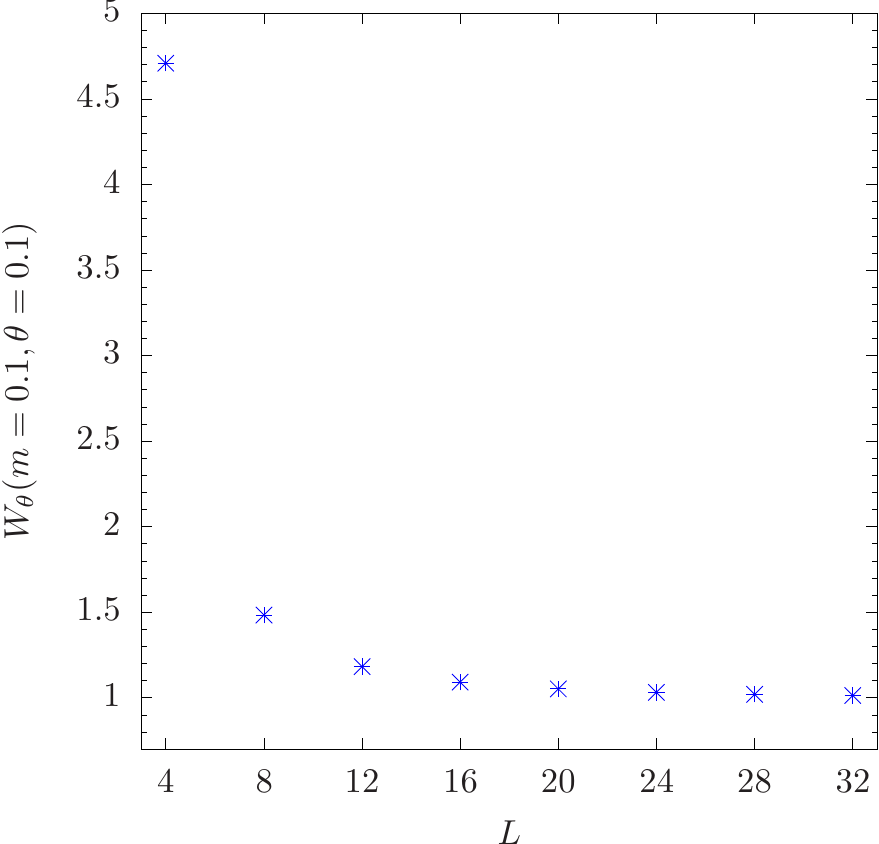}\label{fig:tree_level_test1}}
\hspace{0.4cm}
\subfigure[\emph{Relative variance of the reweighting factor at tree-level, vs $L$, for $\theta = 0.1$.}]{\includegraphics[scale=0.49]{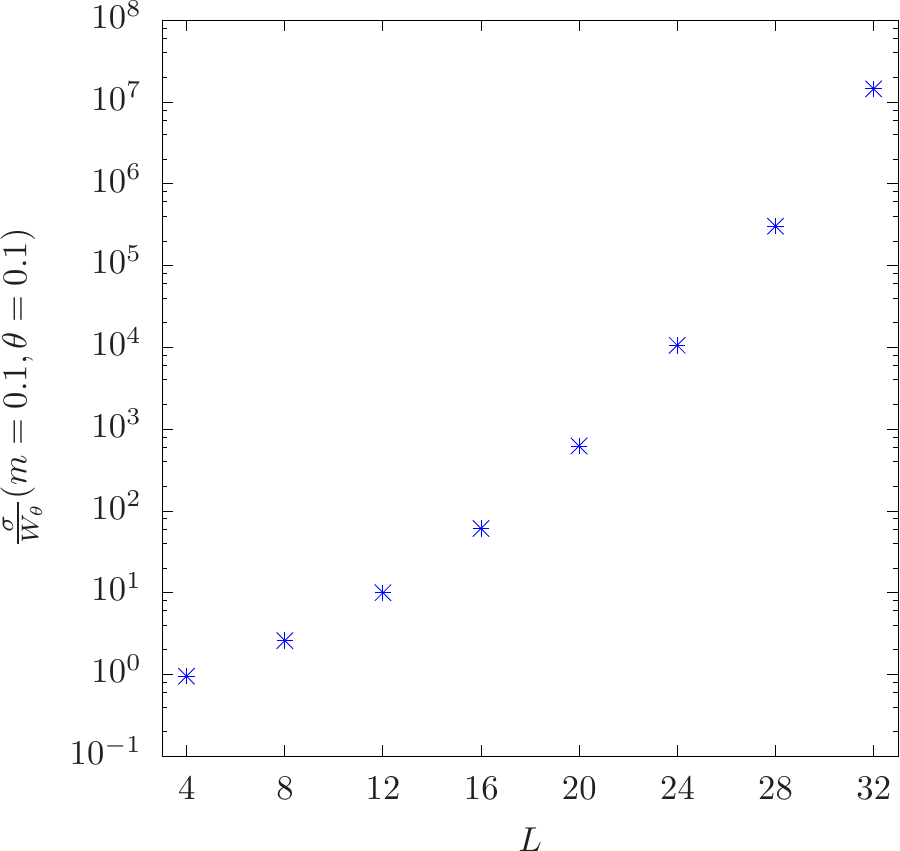}\label{fig:tree_level_test2}}
\hspace{0.4cm}
\subfigure[\emph{Relative variance of the reweighting factor at tree-level, vs $\theta$, for $L=8$.}]{\includegraphics[scale=0.49]{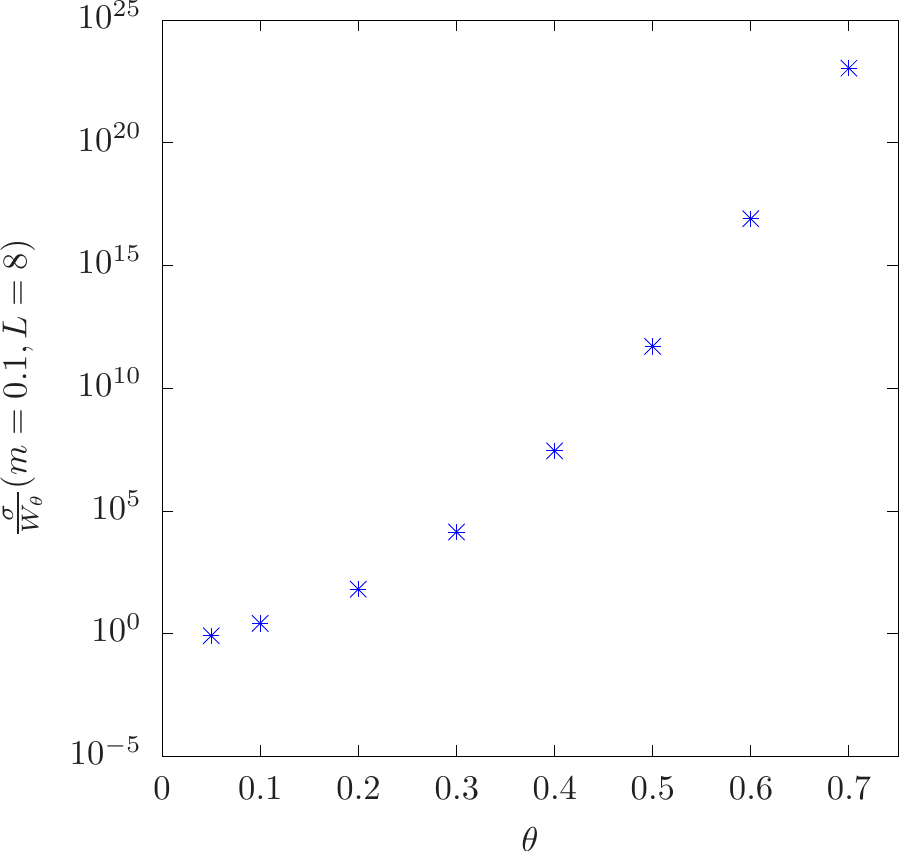}\label{fig:tree_level_test3}}
\caption{\emph{\color{black} Results for the reweighting factor mean and variance \color{black} employing the exact formulae for the tree-level case. Each point corresponds to a cubic lattice of the form $L^4$ with $N_f=2$ and $N_c=2$.}}
\end{center}
\end{figure}

The spectrum of the free Wilson Dirac operator can be easily computed in momentum space, and that provides
insights on the large volume asymptotic scaling of the reweighting factors and their variances.
Given the finite-volume nature of twisting, 
a rather obvious expectation is that the reweighting factors approach the value 1 at fixed $\theta$ and in large volumes,
and that is readily confirmed at tree-level, as shown in Fig.~\ref{fig:tree_level_test1}.
On the other hand the reweighting factor remains an  extensive quantity, and
the corresponding variance grows in the same limit, which is therefore very difficult to be reached numerically.
It is clear from Fig.~\ref{fig:tree_level_test2} that the noise to signal ratio grows at least exponentially as the volume
is increased at fixed $\theta$. The same exponential growth is observed as $\theta$ is made larger at fixed $L$ (see Fig.~\ref{fig:tree_level_test3}).
A factorization of the observable has been proven to be effective in this case in various 
instances~\cite{Luscher:2001up,DellaMorte:2010yp},
and we will pursue a similar approach here, in analogy to what has been done in~\cite{Finkenrath:2013soa} for the case of mass reweighting.
A natural choice is to split the determinants ratio through the following telescopic decomposition
\begin{eqnarray}
	&&  A=	D_W(\theta)D_W^{-1}(0) = \prod_{l=0}^{N-1}A_\ell,\,\text{ with }A_\ell=D_W(\theta_{\ell+1})D_W^{-1}(\theta_\ell)\simeq\mathbf{1}+\ord\left(\delta\theta\right)\;,  \\ 
&&\theta_\ell=\delta\theta \cdot \ell\,\text{ and }\delta\theta=\frac{\theta}{N}\;,
\nonumber
	\end{eqnarray}
where $A_\ell$ are now matrices deviating from the identity 
by small amounts of O$(\delta\theta)$.
The inverse determinant and its error $\varepsilon_{|A^{-1}|}$ are then reconstructed in terms of the $N$ corresponding estimators $1/\det A_\ell$
and $\varepsilon_{|A_\ell^{-1}|}$,
one for each factor $A_\ell$ in the equation above, as 
\begin{equation}
				\frac{1}{\det A} =\prod_{\ell=0}^{N-1}\frac{1}{\det A_\ell}= \prod_{\ell=0}^{N-1}\bigg\langle \frac{\exp\left(-\eta^{(\ell),\dagger} A_\ell \eta^{(\ell)}\right)}{p\left(\eta^{(\ell)}\right)} \bigg\rangle_{p\left(\eta^{(\ell)}\right)},
\end{equation}
and
\begin{equation}
				\varepsilon^2_{|A^{-1}|}  = \sum_{\ell=0}^{N-1}\left[\varepsilon^2_{|A_\ell^{-1}|}\prod_{k\neq\ell}\det \left(A_k\right)^{-2}\right].
\end{equation}
Based on the tree-level results we have presented,
the expression on the r.h.s of the equation above 
is given by a sum of $N$ terms, each one depending exponentially on $\delta \theta$ 
(for values of $\delta \theta$ such that the variance in Fig.~\ref{fig:tree_level_test3} is approximately growing exponentially
with the twisting angle)
and therefore, at
fixed $\delta \theta$, the squared error of the telescopic product is expected to grow linearly with $\theta$. 
That is to be compared to the exponential growth of the error one would obtain by
attempting to compute the ratio of determinants for large shifts in $\theta$ in one single step.

\section{Simulations and results}
For the numerical computations we have used gauge configurations produced for
the $\rm SU(2)$ gauge theory with two fermions in the fundamental representation.
\color{black}
The choice is motivated by the fact that dynamical configurations were locally available  at CP$^3$ from previous and
ongoing projects. The ensembles have been generated using un-improved Wilson fermions and the Wilson plaquette gauge action. 
The model is a QCD-like theory, featuring chiral symmetry breaking and confinement. 
The outcome of the present study should hence remain qualitatively unchanged for the case of lattice QCD.
At tree-level the reweighting factors indeed scale with a power of $N_{\rm c}$, due to the degeneracy
of the eigenvalues of the Dirac matrix.
\color{black}

In Table~\ref{tab:simpar} we collect details about  the ensembles used in this work, for completeness, the value $m_c$ 
of the bare mass parameter yielding massless fermions
is estimated to be $-0.77(2)$ at $\beta = 2.2$ \cite{Lewis:2011zb, Hietanen:2014xca, Arthur:2016dir}.
\begin{table}[htb]
\begin{center}
	\begin{tabular}{|c|c|c|c|c|}
	\hline
	$V$ & $\beta$ & $m$ & $N_\text{cnf}$ & traj. sep.\\
	\hline
	$8^3 \times 16$ & 2.2 & -0.6 & $980$ & 10 \\
	\hline
	$24^3 \times 32$ & 2.2 & -0.65 & $374$ & 20 \\
	$24^3 \times 32$ & 2.2 & -0.72 & $360$ & 10 \\
	\hline
	\end{tabular}
\label{tab:simpar}
\caption{\emph{Ensembles used and simulation parameters}}
\end{center}
\end{table}
We have considered both small and large volumes, which we will discuss separately. 

\begin{figure}[h!t]
\begin{center}
\subfigure{\includegraphics[scale=0.72]{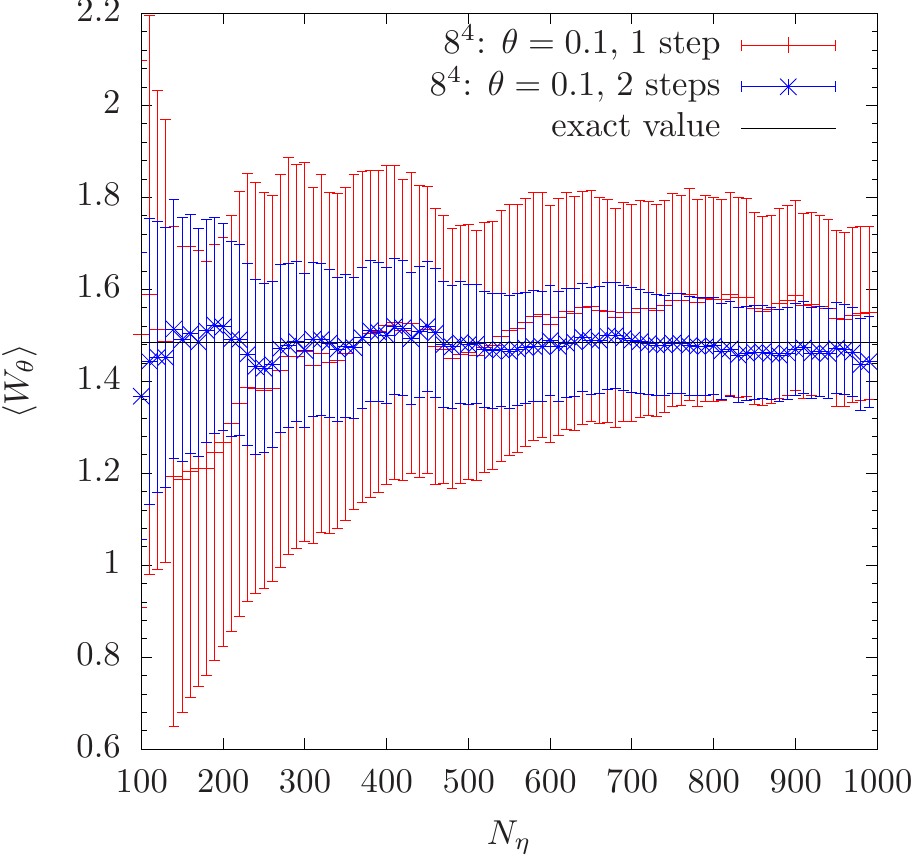}}
\hspace{0.4cm}
\subfigure{\includegraphics[scale=0.72]{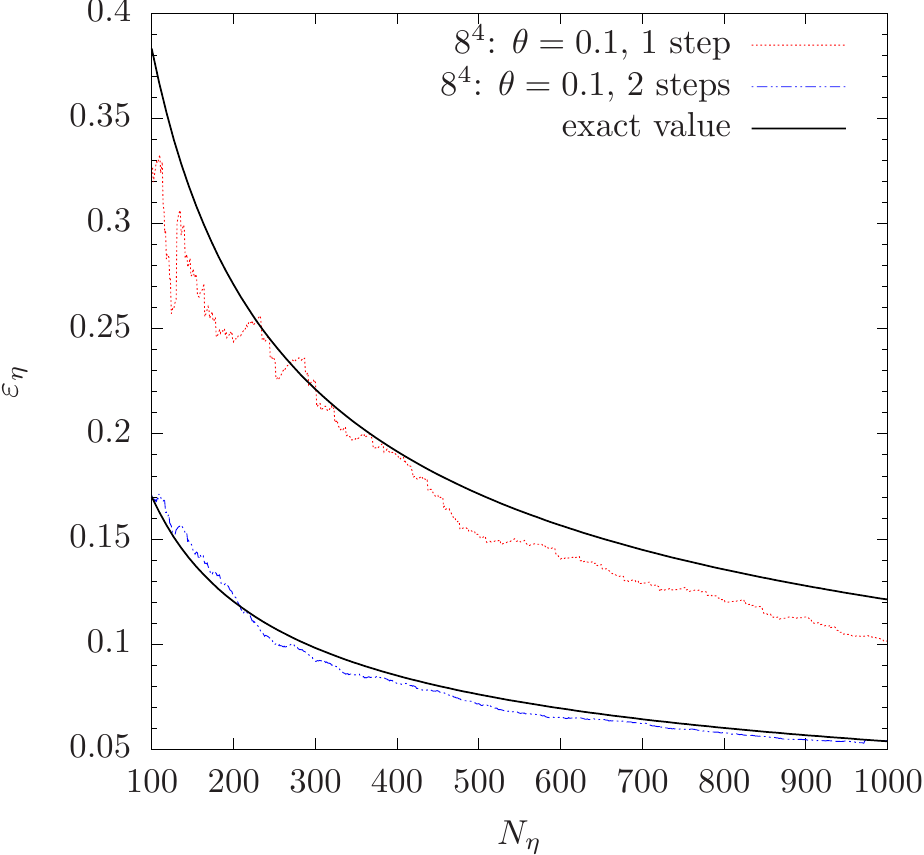}}
\caption{\emph{Comparison between the stochastic and exact estimates of the reweighting factor and its error on the trivial gauge configuration for $\theta=0.1$, $L=8$ using
one and two levels of factorization in the numerical case. Both plotted against the number $N_\eta$ of gaussian vectors in each level.} } 
\label{fig:treecomp}
\end{center}
\end{figure}

The expectation, also from the tree-level studies,  
is that, at fixed values of $\theta$, the effect of twisting and therefore of reweighting is at its largest in 
small volumes, which is also where the stochastic methods should provide reliable estimates.\footnote{In general the relevant parameter
for the effects of twisting is clearly $a\theta/L$ rather than just (the inverse of) $L$. 
That makes the knowledge of the actual lattice extension (in fm) not strictly necessary.}
We restricted ourselves to the case of reweighting to spatially isotropic $\theta$ angles and 
adopted the hermitian $\gamma_5$-version of the Dirac-Wilson operator with two flavors, since that automatically fulfills the applicability condition in 
eq.~\eqref{eq:appl_cond}.\footnote{Notice that $Q[U,0]$ and $Q[\tilde{U},0]$ commute.}
As a test of the stochastic method and in order to get an idea about the number $N_\eta$ of gaussian vectors necessary to obtain a reliable estimate
of the reweighting factor, we computed it on the trivial ($U_\mu(x)={\mathbb{I}}$) gauge configuration and compared it to the analytical tree-level prediction.
The results are shown in Fig.~\ref{fig:treecomp}.
%
There, by looking at the statistical error, one sees that about 300 gaussian vectors are necessary for $\varepsilon_\eta$ to reach the correct scaling with $N_\eta$.

\begin{figure}[h!t]
\begin{center}
\subfigure{\includegraphics[scale=0.72]{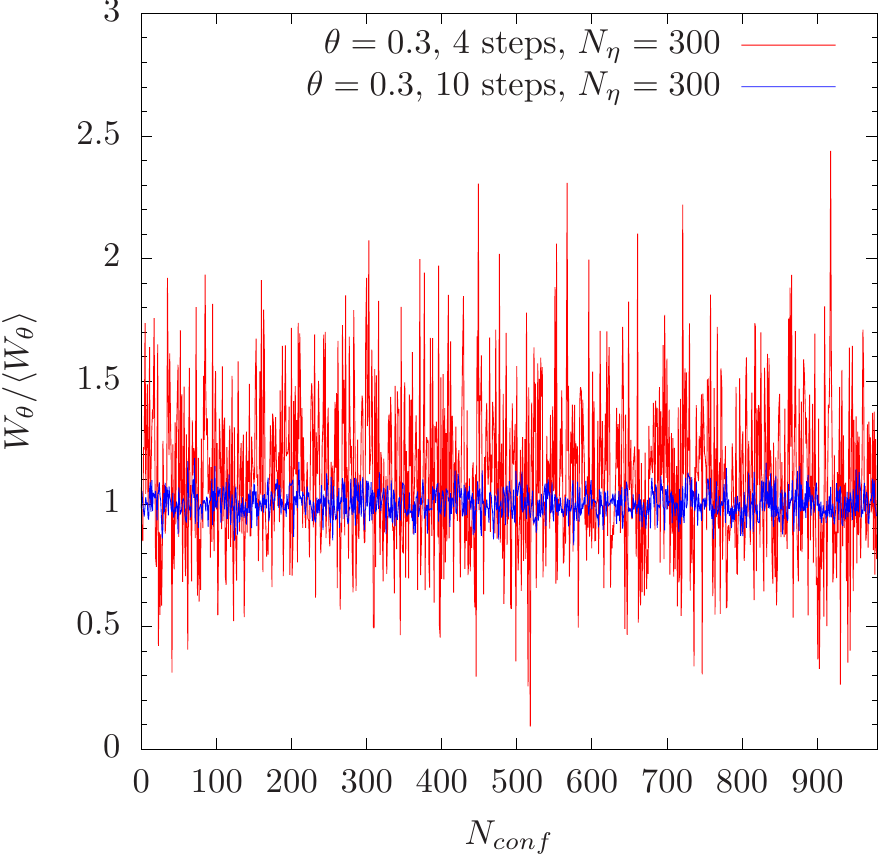}}
\hspace{0.4cm}
\subfigure{\includegraphics[scale=0.72]{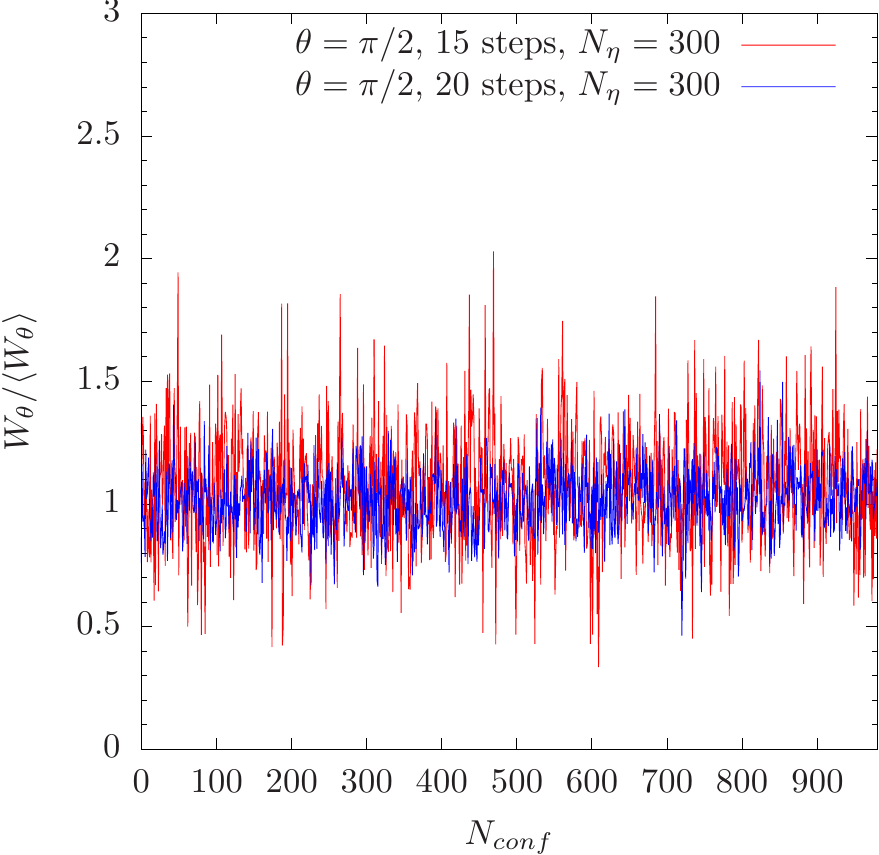}}
\caption{\emph{Monte Carlo history of the reweighting factor relative to its mean for the $8^3 \times 16$ ensemble and with $N_\eta$=300.}}
\label{fig:maybe}
\end{center}
\end{figure}

\begin{figure}[h!t]
\begin{center}
\subfigure[\emph{Mean of the reweighting factor for $\theta = 0.3$ as a function of $N_\eta$.}]{\includegraphics[scale=0.72]{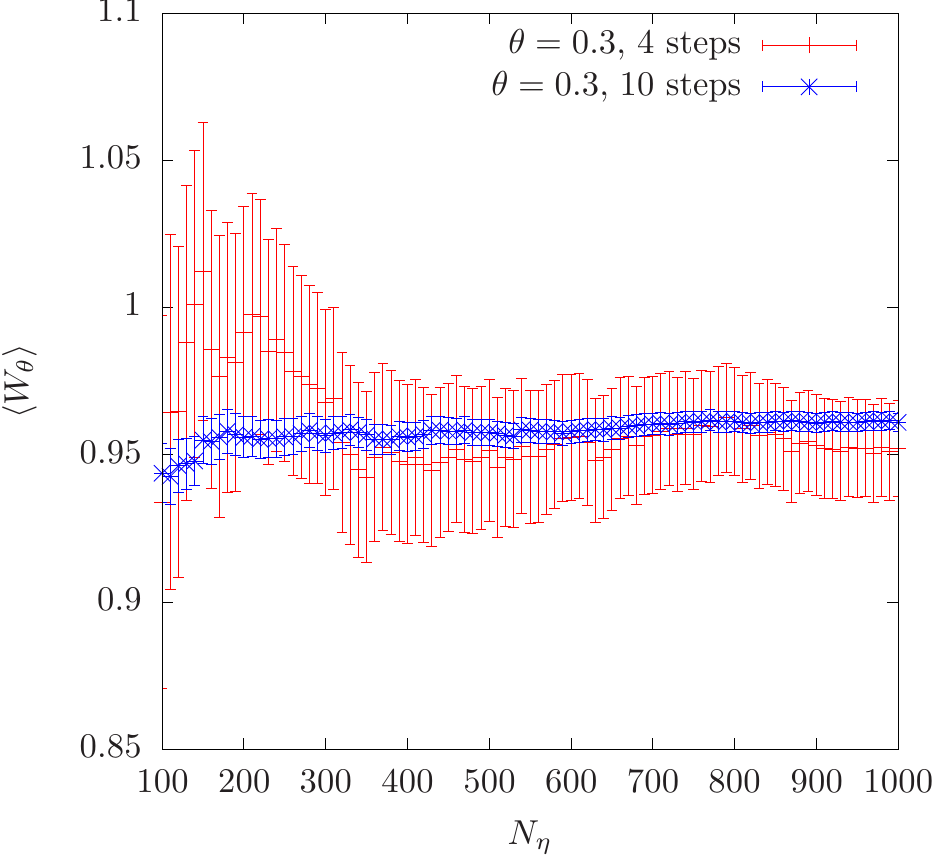}\label{fig:mc_hist_1}}
\hspace{0.4cm}
\subfigure[\emph{Mean of the reweighting factor for $\theta = \pi /2$ as a function of $N_\eta$.}]{\includegraphics[scale=0.72]{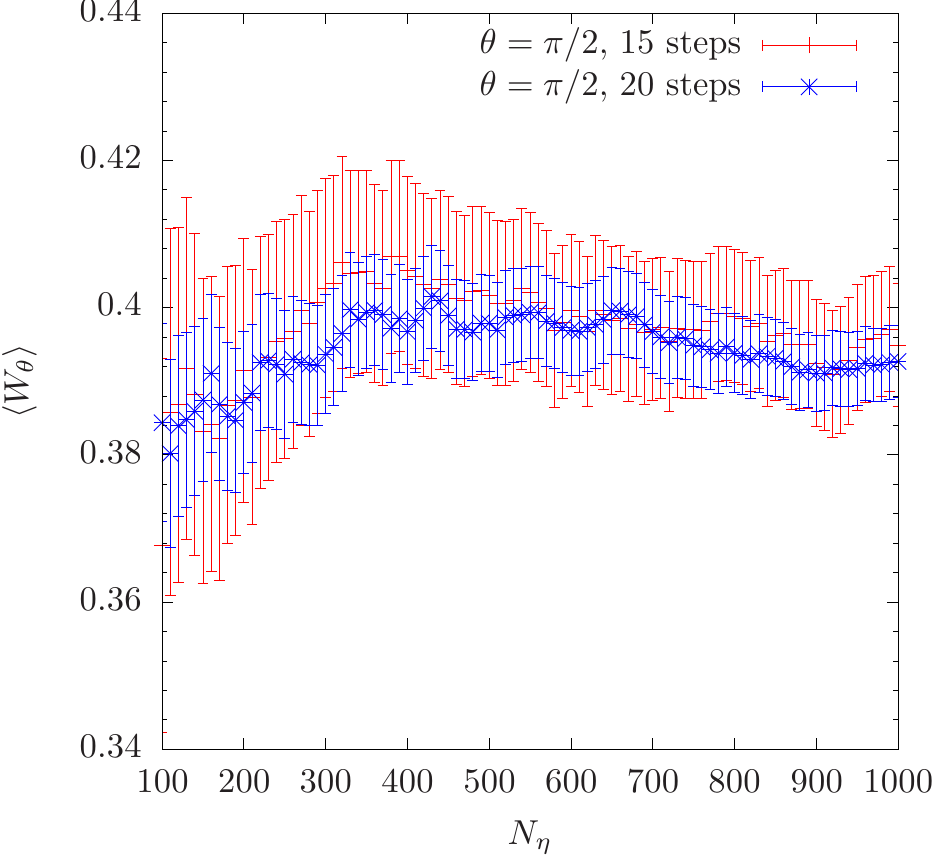}\label{fig:mc_hist_2}}
\caption{\emph{Monte Carlo \color{black} average \color{black} of the reweighting factor, over the entire number of configurations
\color{black} vs $N_\eta$ \color{black}. The figures correspond to a volume $V = 8^3 \times 16$.}}

\end{center}
\end{figure}

Turning now to actual Monte Carlo data, in particular from the $8^3 \times 16$ ensemble, we found that under a similar  condition ($N_\eta \gtrsim 300$)
the reweighting factor on each configuration \color{black} deviates by more than a factor 10 in magnitude from \color{black}
its gauge average in a few cases only (see Fig.~\ref{fig:maybe}).
This prevents the averages to be dominated by ``spikes'', which would cause large statistical fluctuations, and sets a lower limit on $N_\eta$. 
In Figs.~\ref{fig:mc_hist_1} and~\ref{fig:mc_hist_2} we show the mean of the reweighting factor, with each point resulting from an average over 1000 configurations, as a function of $N_\eta$.
A good scaling of the error is  visible, according to $N_\eta^{-1/2}$, up to $N_\eta \approx 500$.
At that point the statistical noise saturates the gauge noise, and therefore the error on the gauge average does not decrease any further by increasing $N_\eta$.
This sets an upper limit to about 600 for the number of gaussian vectors to be used in the stochastic evaluation of the determinants ratio.
\color{black}
In addition, it imples that for $N_\eta \gtrsim 600$ one can safely consider the gauge noise only in the error analysis.
In the following, we do that by a standard jackknife plus binning procedure.
\color{black}

\subsection{Small volumes}

In the small volume regime the effect of partial twisting and the associated breaking of unitarity may be large.
In order to isolate the contribution due to the determinants ratio, we looked first at the plaquette, for which
the entire dependence on $\theta$ obviously comes from the quarks in the sea only.
Here and in the following we neglect autocorrelations since measurements are separated by 10 to 20 molecular dynamics units. 
In any case, a binning procedure, using bins of length up to $10$, provides entirely consistent results.

\begin{figure}[h!t]
\begin{center}
\subfigure[\emph{\color{black} Average \color{black} plaquette after reweighting only one flavor.}]{\includegraphics[scale=0.72]{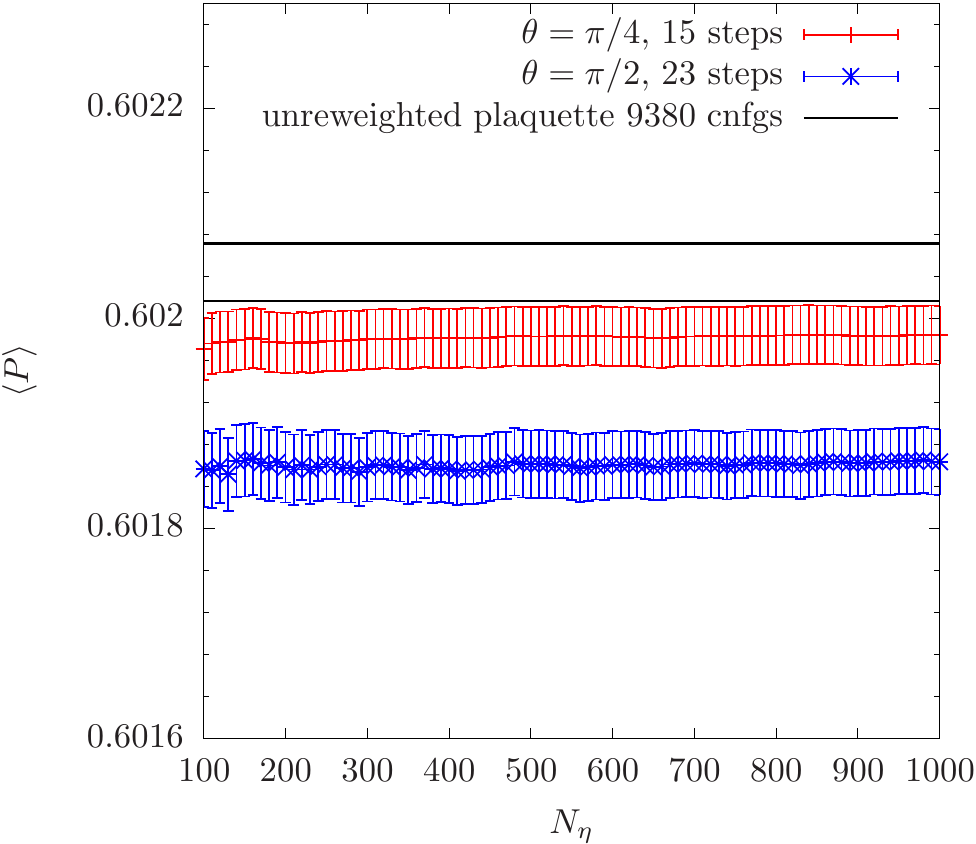}\label{fig:plaquette_small_1}}
\hspace{0.4cm}
\subfigure[\emph{\color{black} Average \color{black} plaquette with both flavors reweighted.}]{\includegraphics[scale=0.72]{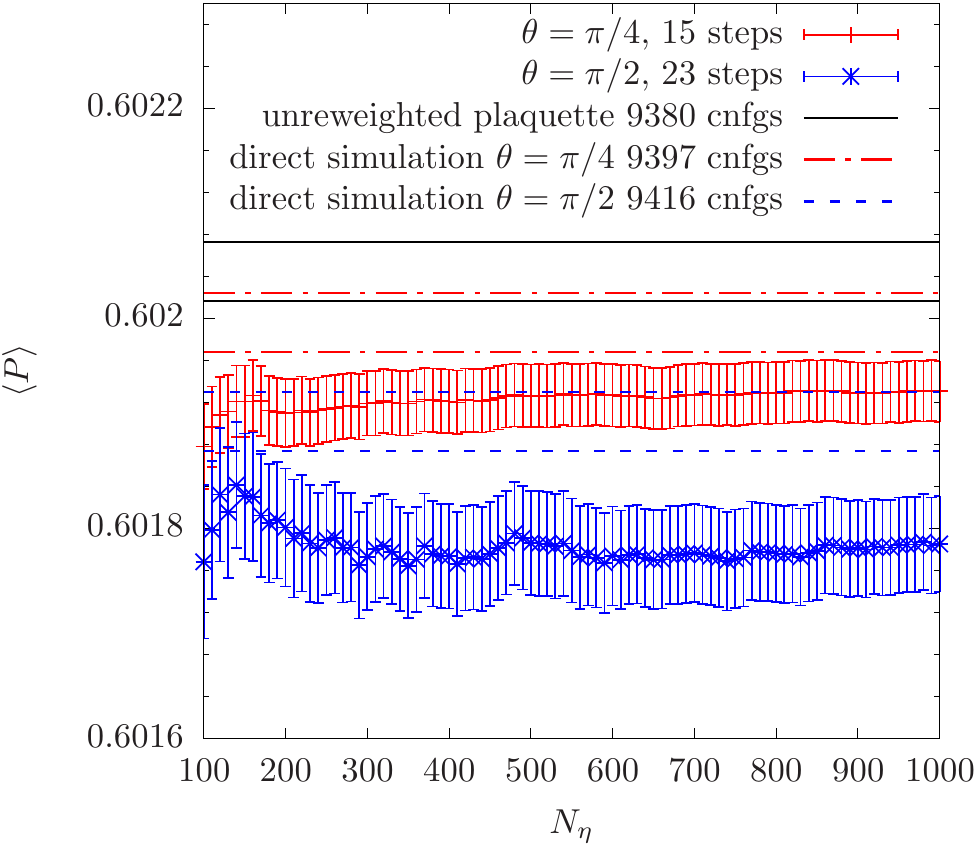}\label{fig:plaquette_small_2}}
\caption{\emph{Monte Carlo history of the reweighted plaquette. The figures correspond to a volume $8^3 \times 16$.}}
\end{center}
\end{figure}

In Fig.~\ref{fig:plaquette_small_1} we show the results after reweighting only one flavor, i.e., by taking the square root of the stochastic
evaluation of the determinants ratio estimated for the $Q^2[U, \theta]$ operators~\cite{Aoki:2012st,Finkenrath:2012cz}.
Notice that here and in the following, whenever the root-trick above is used, we consider rather heavy quarks and pions ($am_\pi \geq 0.45$) and 
we therefore do not expect ambiguities in the sign of the one-flavor determinant.
Effects are visible within statistical errors for large values of $\theta$ only.
Those are more pronounced when both flavors are reweighted, as depicted 
in Fig.~\ref{fig:plaquette_small_2}. In addition, in this case, the reweighted 
results can be checked by a direct HMC simulation\footnote{For the
one-flavor case one would have to consider the RHMC algorithm.} at $\theta \neq 0$. Notice that we are discussing permil shifts, which we access by 
using very large statistics ($\approx$ 10000 configurations) for such a cheap
quantity as the plaquette.
That changes from $\theta=0$ to $\theta=\pi/2$ by about 3.5 combined sigmas and
the result is reproduced by reweighting from $\theta=0$ within two combined 
statistical deviations (for $N_\eta \gtrsim 600$).
\color{black}
We interpret this slight tension as signalling the limit of validity of the
reweighting method for the application discussed here. In the following, we therefore restrict the values
of the twisting angle to the interval $[0,\, \pi/2]$.
\color{black}
%

\begin{figure}[h!t]
\begin{center}
\includegraphics[scale=0.72]{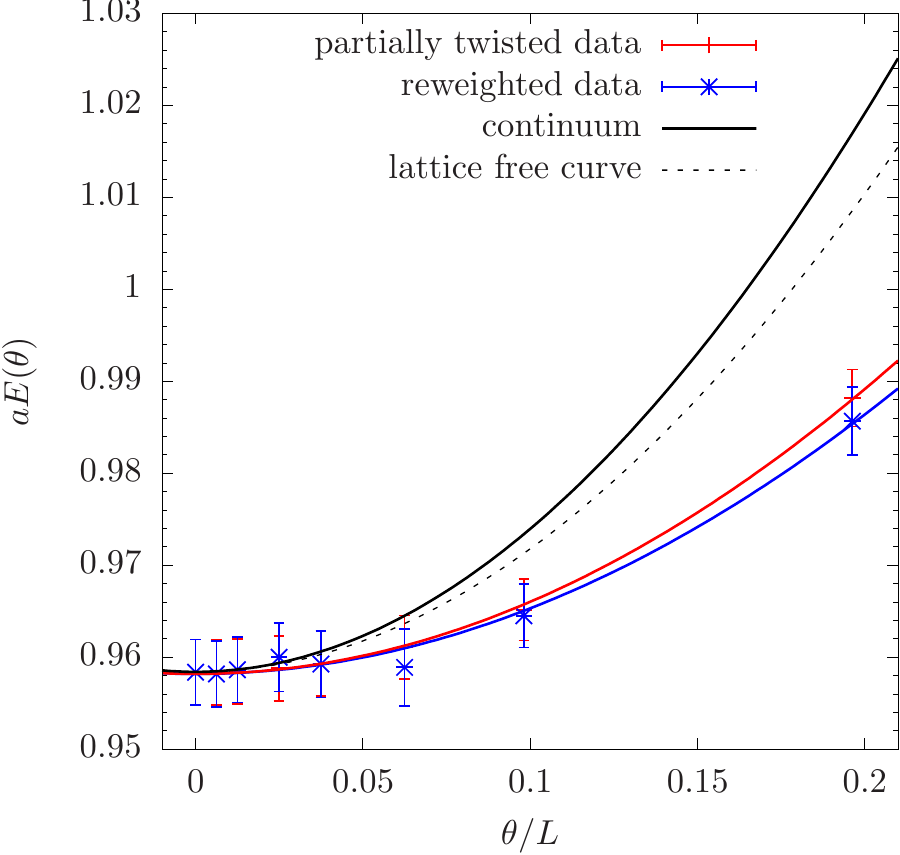}
\caption{\emph{Pion dispersion relation for $V = 8^3 \times 16$. Each point is obtained with a different (growing with $\theta$) number of independent steps in the determination of the reweighting factor and $N_\eta \gtrsim 600$ in each step.}}
\label{fig:disp_rel_small}
\end{center}
\end{figure}

The other quantity  we have analyzed is the pion dispersion relation.
After twisting only one flavor in the valence, the lowest energy state coupled
to a spatially summed interpolating field
is expected to become a ``quenched'' pion with momentum $\vec{p} = \pm\vec{\theta} / L$. 
In order to remove this quenching effect we have reweighted the relevant correlators 
for the twisting of one flavor in the sea and we have extracted the effective energies
from their time-symmetrized versions.
In Fig.~\ref{fig:disp_rel_small} we display the results for the dispersion relation and we compare them to 
the un-reweighted, partially twisted, data (i.e., with twisting in the valence only), to the continuum prediction $(a E)^2 = (a m_\pi)^2 + 3 \theta^2 / L^2$ 
(we use $\vec{\theta}=\theta(1,1,1)$)
and to the lattice free boson theory prediction $\cosh(a E) = 3 + \cosh(a m_\pi) + 3 \cos(\theta / L)$.
Over the entire range of  $\theta$ values explored there is no significant effect within errors. 
We will discuss possible explanations of the
discrepancy between the reweighted data and the lattice free prediction 
for large twisting angles, in the next Section.
Let us remark that, following~\cite{Boyle:2008rh}, all two-point functions 
have been computed using $Z_2 \times Z_2$ single time-slice stochastic sources.
	\subsection{Large volumes}
\begin{figure}[h!t]
\begin{center}
\subfigure[\emph{``Heavy'' case of $m\simeq -0.65$.}]{\includegraphics[scale=0.72]{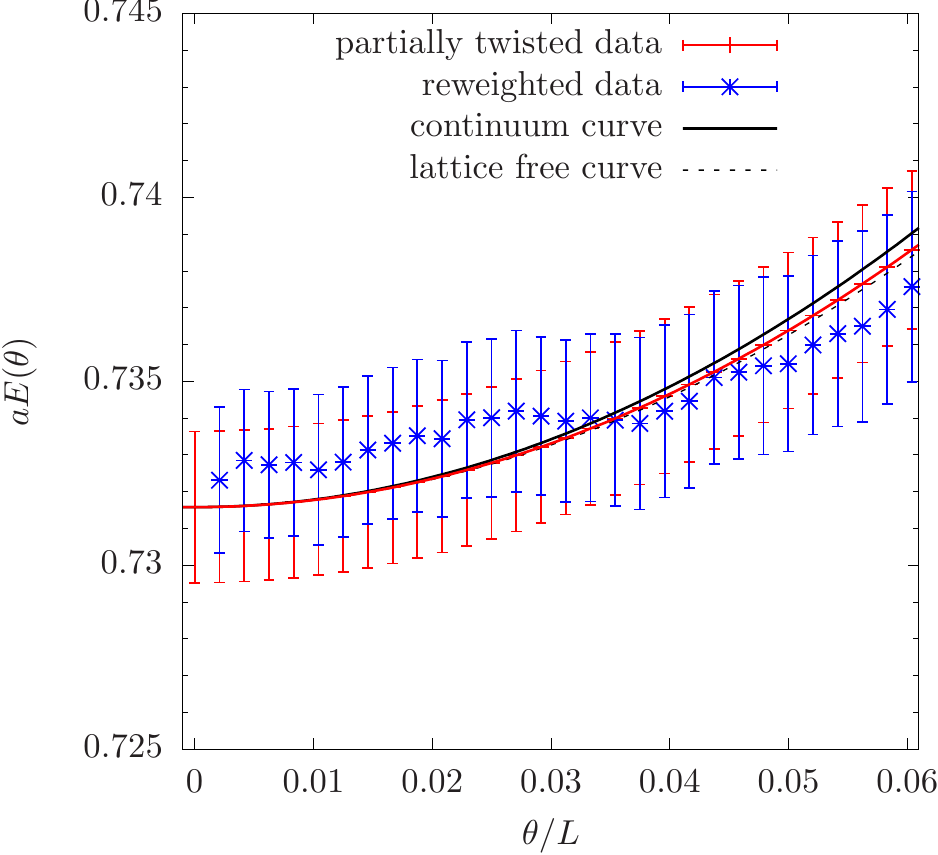}}
\hspace{0.4cm}
\subfigure[\emph{``Light'' case of $m\simeq -0.72$.}]{\includegraphics[scale=0.72]{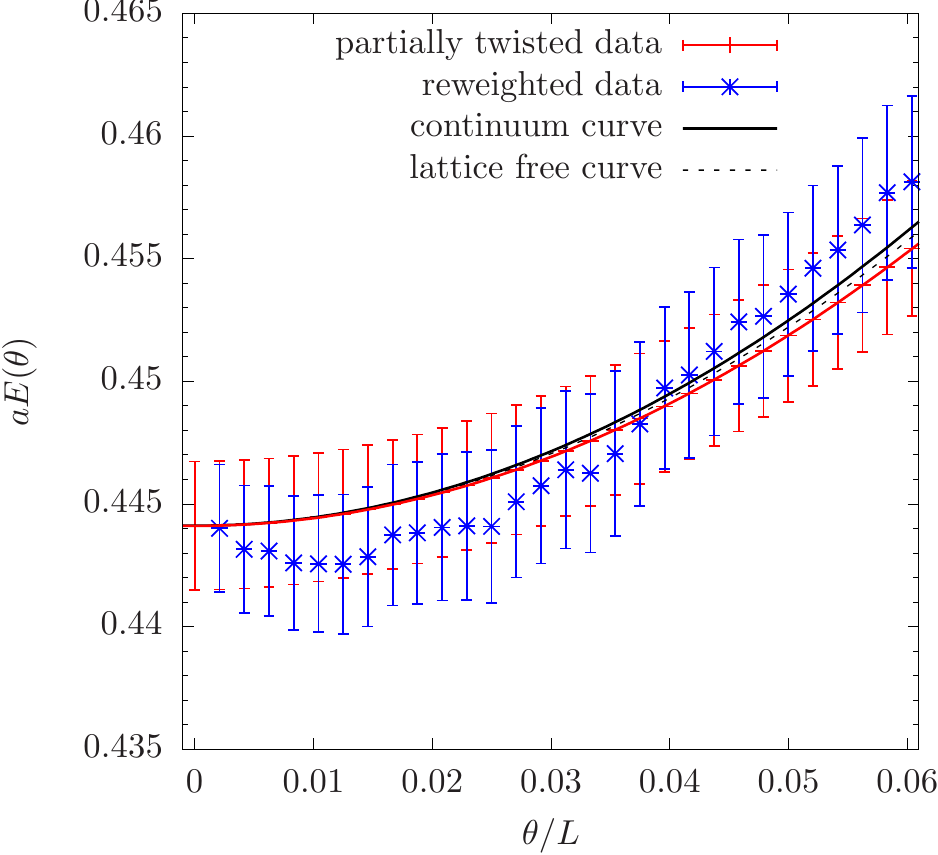}}
\caption{\emph{Pion dispersion relation for $V = 24^3 \times 32$. The reweighting factor at a given $\theta$ is obtained by a telescopic product involving all the previous ones, each one estimated using $N_\eta \gtrsim 600$.}}
\label{fig:disp_rel_large}
\end{center}
\end{figure}

As suggested by the tree-level studies, in large volumes, the accuracy in the determination of the reweighting factors and the overall effect of twisting
are very much reduced, compared to the previous case.
We have looked at the pion dispersion relation for two different values of $m_\pi$ as 
we expect to detect possibly sizeable effects for rather light quarks.
However, at the volume considered ($V=24^3 \times 32$), it appears as one can safely neglect any breaking of unitarity.
The results are shown in Fig.~\ref{fig:disp_rel_large}.
Indeed, reweighting does not seem to yield any significant effect within the half a percent statistical errors. 

\section{Quark mass dependence on $\theta$}

As mentioned above, non-periodic boundary conditions ($\theta\neq 0$) for fermionic fields are equivalent to introducing a constant $\rm U(1)$ interaction, 
through an external field $B_\mu(x)=B_\mu$ coupled to periodic fermions. That amounts to replacing the links $U_\mu$ with the links $\tilde{U}_\mu$ given by
\begin{align}
\tilde{U}_\mu(x) = \e^{iaB_\mu}U_\mu(x)\;,
\end{align}
where $B_\mu = \theta_\mu / L_\mu$.
In this case the Partially Conserved Axial Current (PCAC) relations remain unaffected since the new links $\tilde{U}_\mu(x)$, being proportional to the 
identity in flavor, still commute with the Pauli matrices. That implies that the vector transformations are still exact symmetries at finite lattice spacing with
Wilson fermions, and that the PCAC relation remains formally the same.
Cutoff effects on the other hand depend on the choice of boundary conditions, that has actually been exploited in order to compute improvement 
coefficients (see, for example, Refs.~\cite{Luscher:1996ug,Durr:2003nc}), within the Symanzik improvement programme for Wilson fermions in QCD.
In some instances, even after improvement, the quark mass defined through the PCAC relation in rather small volumes, turned out
to have a quite pronounced residual (i.e., O$(a^2)$) dependence on the boundary conditions~\cite{Sommer:2003ne}. That is expected
to be even larger here, with un-improved Wilson fermions. For completeness, the bare PCAC quark mass $m$ can be defined through the spatially integrated axial Ward identity as:
\begin{equation}
m_{\rm PCAC}= \frac{\partial_0\langle A_0(x_0) O(0)\rangle}{2\langle P(x_0) O(0) \rangle}\;,
\end{equation}
with $A_\mu$ the axial current, $P$ the pseudoscalar density (both spatially summed over the $x_0^{\rm th}$ time-slice) 
and $O$ an interpolating field, which in our case will be simply given by the pseudoscalar density localized at the origin.

For the present discussion, it is useful to introduce $\theta_{\rm v}$ as the spatially isotropic twisting angle used for all fermions
in the valence and  $\theta_{\rm s}$ as the corresponding one for all fermions in the sea. We produced small volume configurations
for different values of $\theta_{\rm s}$ and on those we measured quark and pion masses while changing $\theta_{\rm v}$.
That can actually be used to get an idea of what the effect of reweighting should be, and we will
see that the observations above are confirmed. In particular, at fixed  $\theta_{\rm v}$, results depend 
at most at the percent level only on  $\theta_{\rm s}$.

\begin{figure}[h!t]
\begin{center}
\subfigure{\includegraphics[scale=0.72]{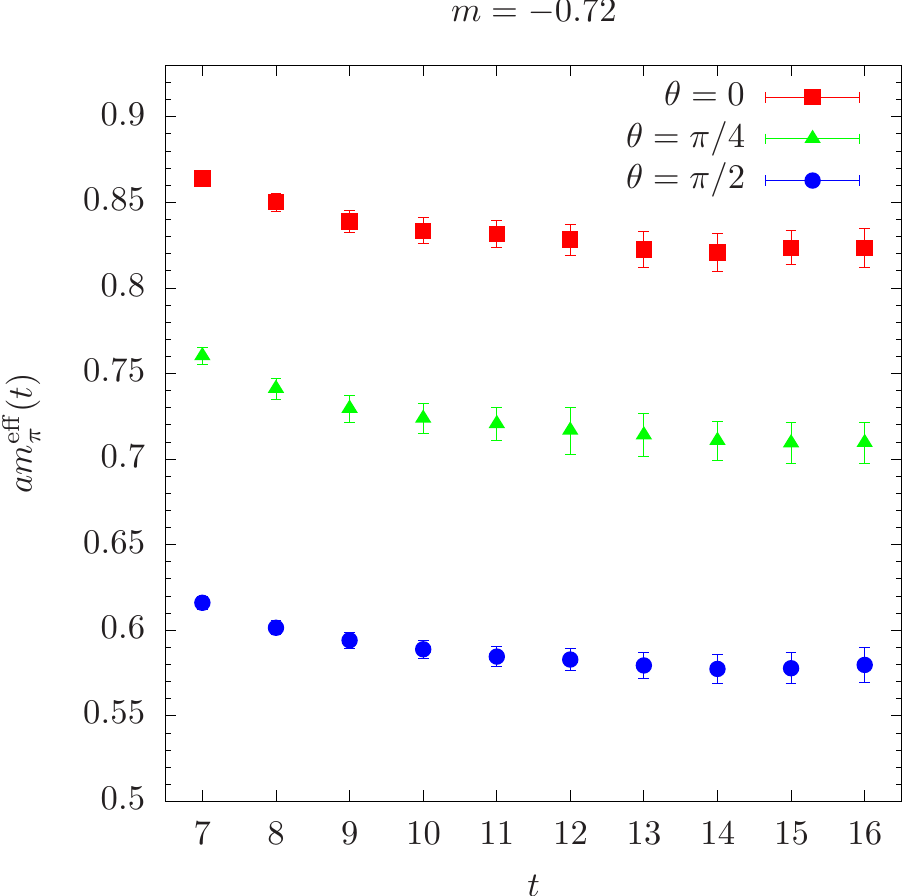}}
\hspace{0.4cm}
\subfigure{\includegraphics[scale=0.72]{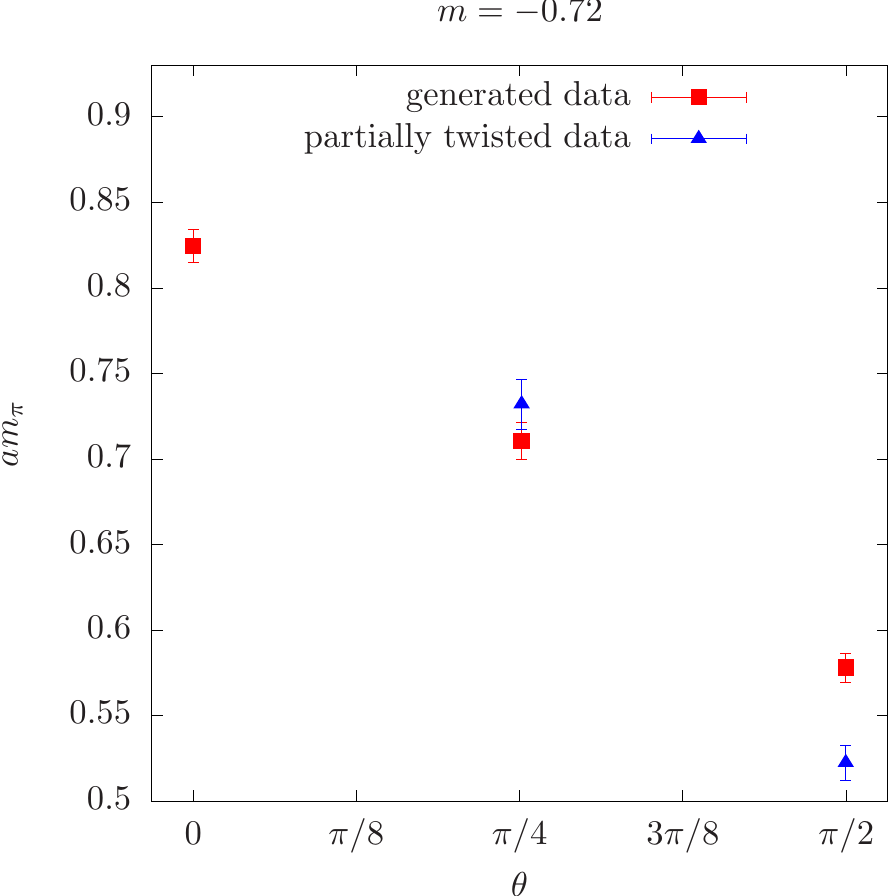}}
\caption{\emph{Pion effective masses for $\theta_{\rm s}=\theta_{\rm v}=\theta$ on a $8^3 \times 32$ lattice at $\beta=2.2$, $m=-0.72$ (left panel) and comparison
of the corresponding plateau masses (red squares, right panel) with the results from $\theta_{\rm s}=0$ as a function of $\theta=\theta_{\rm v}$ 
(blue triangles, right panel).}
}
\label{fig:thetaseandval}
\end{center}
\end{figure}

In Fig.~\ref{fig:thetaseandval} we show the pion effective masses computed on about 4000 independent configurations generated for lattices 
of size $8^3 \times 32$ at $\beta=2.2$ and $m=-0.72$. All the results in the left panel refer to unitary points, i.~e., 
with $\theta_{\rm s}=\theta_{\rm v}$.
Naively one would expect the results to lie on top of each other as they all correspond to zero-momentum pions at the same bare parameters.
We ascribe the difference to the dependence of the critical bare mass
$m_c$ on both $\theta_{\rm s}$ and $\theta_{\rm v}$. Since $m_c$ is obtained from the PCAC operator identity the dependence
on the twisting angles is a boundary, therefore finite volume, cutoff effect.
This explains why the continuum dispersion relation is rather poorly reproduced in small volumes and for large values
of the twisting angle, as shown in Fig.~\ref{fig:disp_rel_small}. Upon twisting, not only the pions get boosted, but also at least
one of the quark  masses decreases, such that the two effects partly compensate. 
In the right panel of Fig.~\ref{fig:thetaseandval} the final estimates of the masses are shown as a function of $\theta=\theta_{\rm v}$.
The square points are the unitary ones, corresponding to the plateaux in the left panel, whereas the
triangle ones are obtained by using configurations produced at $\theta_{\rm s}=0$ on which
two-point functions are computed for different values of $\theta_{\rm v}=\theta$. 
It is clear that by reweighting the sea twisting angle to the valence one a percent effect at most could have been produced here 
in the case  $\theta_{\rm s}=0$, $\theta_{\rm v}=\pi/2$.

\begin{figure}[h!t]
\begin{center}
\includegraphics[scale=0.8]{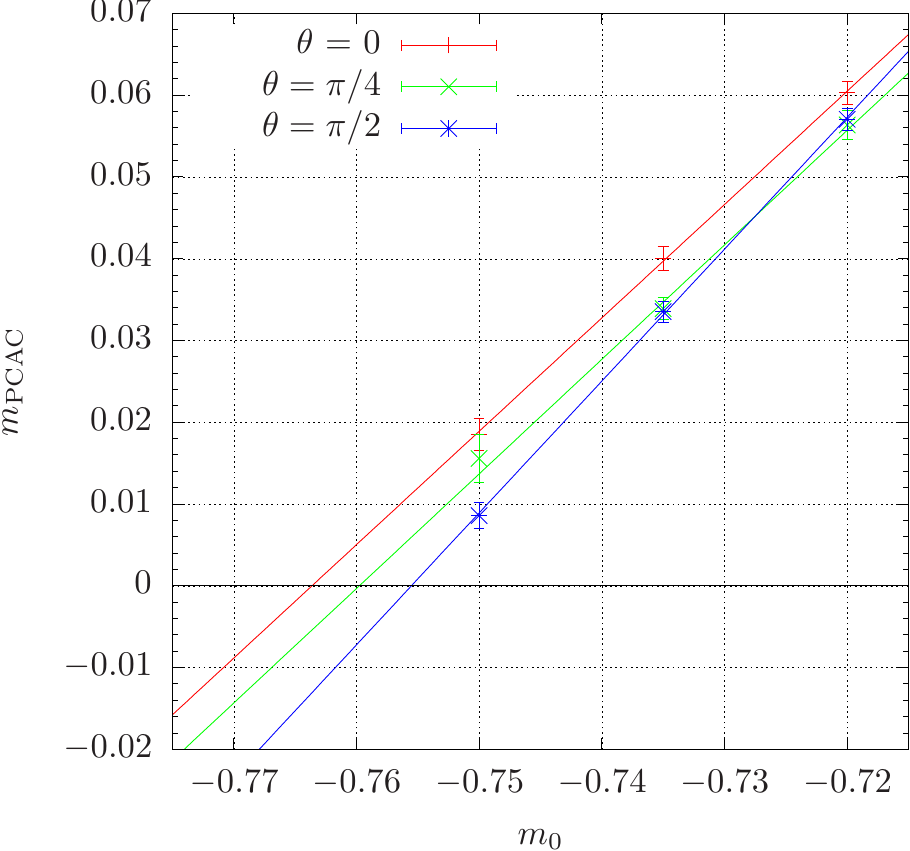}
\caption{\emph{$m_{\rm PCAC}$ as a function of $m$ computed for $m=-0.72$, $-0.735$ and $-0.75$ at $\beta=2.2$ and $V=8^3 \times 16$.
The curves correspond to three different values of $\theta=\theta_{\rm v}=\theta_{\rm s}$.}}
\label{fig:mpcac-vs-m0}
\end{center}
\end{figure}

The bare PCAC quark mass, in the un-improved theory, can be related to the bare parameter $m$ as
\begin{equation} 
m_{\rm PCAC}(\beta, \theta)=Z(\beta,\theta)\left( m-m_c(\beta,\theta) \right)\; ,
\end{equation}
where $Z$ is a normalization factor and $m_c$ is the value of the bare mass parameter defining the massless limit. 
As a consequence of the breaking of chiral symmetry with Wilson fermions $m_c$ is different from zero, as opposite
to the case of Ginsparg-Wilson or staggered fermions, where at least some axial transformations are preserved 
and that is enough to rule out an additive renormalization of the bare mass.
We are here restricting the attention to the case $\theta_{\rm s}=\theta_{\rm v}=\theta$ and we are working at fixed $\beta$.
By looking at $m_{\rm PCAC}$ as a function of $m$, for $m$ slightly larger than $m_c$, one obtains a family of linear
curves parameterized by $\theta$. The slope of the curves is given by $Z$, while the value of $m$ where the curves
intercept the horizontal axis corresponds to $m_c$. That is exactly what is shown in Fig.~\ref{fig:mpcac-vs-m0} for 
$m=-0.72$, $-0.735$ and $-0.75$ (with $\beta$ fixed to $2.2$ and $V=8^3 \times 32$). Whereas the dependence 
of $Z$ on $\theta$ is not significant, $m_c$, as anticipated, changes substantially with the twisting angle.
At $\theta =  \pi/2$ and for $m=-0.75$ the PCAC mass is roughly one half of the value at $\theta = 0$. 
\color{black} We find that \color{black} the corresponding ratio for pion masses is within $1.5$ and $1.7$, which, given the small volumes we have 
considered, is quite consistent with the scaling in a chirally broken theory.
%

Finally, let us remark that
similar shifts in the quark mass have been discussed for the case of constant external magnetic fields~\cite{Bali:2015vua}.
We emphasize that here we are showing that such effects are present also in the case of vanishing magnetic field and constant (vector) potential.

\vspace{-0.3cm}
\section{Conclusions}
We explored in detail an application of reweighting techniques to the case of modifications in the spatial periodicity of fermions.
We have thoroughly studied the approach at tree-level and we have provided constraints for the convergence 
of the stochastic estimates of all the gaussian moments of the reweighting factors. We have also
established the large volume scaling of the average and the variance at tree-level.

At the numerical level we have performed a complete and detailed study of purely gluonic as well as fermionic quantities in both large
and small volumes. We considered the plaquette and the pion dispersion relation, the latter in the two regimes.
In both cases we found the effects of reweighting to be at the sub-percent level for values of $\theta$ up to $\pi/2$.
\color{black}
In our implementation, for the large volumes considered, we found that the reweighting method, with $N_\eta \approx 600$ 
is a factor 4-5 computationally cheaper compared to generating new configurations for different values of the twisting angle,
assuming that about 20 molecular dynamics units are needed to decorrelate subsequent configurations.
We performed this comparison in units of matrix-vector multiplications.
\color{black}

Perhaps the most important observation, which, as far as we know, has so far not been investigated in a dedicated way in the literature,
is on the dependence of the critical mass for Wilson fermions on the periodicity phases in the boundary conditions.
Although a cutoff effect, it could be rather large in a theory with O$(a)$ discretization effects. 
Since the result is that the hadron masses, and not only their momentum, change as $\theta$ changes, the 
corresponding dispersion relations may look very different at finite and coarse lattice spacings compared to the
continuum predictions. 

\vskip 0.2cm

\noindent
{\bf Acknowledgements.}
This work was supported
by the Danish National Research Foundation DNRF:90
grant and by a Lundbeck Foundation Fellowship grant.

\end{document}